\documentstyle[aas2pp4]{article}
\begin{document}
\title{\bf Cluster Mergers as Triggers of Star Formation and Radio
Emission: A Comparative Study of the Rich Clusters A2125 and A2645}

\author{Frazer N. Owen$^1$}
\affil{National Radio Astronomy Observatory$^2$ \\ 
Socorro, New Mexico  87801}

\author{Michael J. Ledlow$^1$}
\affil{University of New Mexico, Dept. of Physics \& Astronomy\\Albuquerque, NM
87131}

\author{William C. Keel$^1$}
\affil{University of Alabama, Dept. of Physics and Astronomy\\ Tuscaloosa, AL}

\and

\author{Glenn E. Morrison$^1$,$^3$}
\affil{University of New Mexico, Dept. of Physics \& Astronomy\\Albuquerque, NM
87131}
$^1$Visiting Astronomer, Kitt Peak National Observatory, National Optical 
Astronomy Observatories, operated by the Association of Universities for 
Research in Astronomy, Inc., under contract with the National Science 
Foundation. 

$^2$The National Radio
Astronomy Observatory is operated by  Associated Universities, Inc., under
a cooperative agreement with the National Science Foundation.

$^3$ Also National Radio Astronomy Observatory, P. O. Box O, Socorro, NM 87801.
\baselineskip 24pt 

\begin{abstract} 

	We report a detailed optical study of the clusters Abell 2125 and
2645. These clusters are very similar in redshift ($z \approx 0.25$) and 
richness (Abell class 4), yet contrast strongly in blue fraction and
radio-galaxy populations. \cite{bo84} report that A2125 and A2645 have 
blue-galaxy fractions of 0.19 and 0.03 respectively, while
radio observations with the VLA and subsequent optical identifications
on the digital Palomar Sky Survey show an apparent
excess of radio galaxies in A2125 relative to A2645 (\cite{dw96}).

	Our spectroscopic observations confirm this difference. We find
27 radio galaxies to be members of A2125, and only four in A2645, based
on (nearly) complete observations to the same limiting magnitude and radio flux
density. The radio galaxies in A2125 extend over about 5 Mpc (assuming
$H_0=75$ km s$^{-1}$ Mpc$^{-1}$) along
a band running from NE to SW of the cluster center. About half the
radio galaxies are red and have optical spectra which resemble old
stellar populations. The other half are blue with emission lines, most
of which indicate an origin in star formation rather than AGN. 
Many of the blue galaxies are located in a distinct clump located about 2 
Mpc in projection from the
cluster center.

	The excess population of radio galaxies in A2125 occurs entirely
at radio luminosities $< 10^{23}$ W Hz$^{-1}$, where one expects
star-formation to be primarily responsible for the radio emission.
Most of these radio galaxies have optical properties most consistent
with systems later than E/S0. However, the optical line luminosities
are often weaker than one would expect for the star formation rates
implied by the radio emission. Thus we suspect that dust obscuration,
larger than is usually found locally, hides most of the star-forming
regions optically.

	The existence of a cluster-cluster merger in progress in A2125
seems likely to play some role in these phenomena, although the details
are obscure. 
 

\end{abstract}

\clearpage 

\section{Introduction}
\baselineskip=0.5\baselineskip
	One of the strongest observational signatures of galaxy evolution
is the Butcher-Oemler (BO) effect (\cite{bo84}), an increase in the 
fraction of blue galaxies found in rich clusters with increasing redshift. 
This largely indicates that star formation was more active in these
clusters in the past than it is now. However, the BO definition
of ``blue" is relative to the populous red cluster members, objects whose
colors indicate little star formation within the $\sim 10^9$ years
before the epoch of observation. Most studies of the BO effect in clusters
have concentrated on the cores, for the practical reason that
foreground or background contamination is least important and most
easily corrected there. However, many of the proposed mechanisms to mediate the BO
effect, and hence the overall evolution of galaxies in the cluster
environment, have effects spread across the cluster, and in some cases
predict different behavior far from the core.
Thus if one could study cluster populations further from the
cluster center as well as closer to the epoch of the star formation, one might
expect to learn more about the origins of the BO effect.

	Tracing star formation might be done most ideally via far-infrared
mapping, since some star formation can be hidden by surrounding dust. This
remains difficult due to instrumental field-of-view and sensitivity
limitations, but the well-established relation between far-IR and centimeter
radio fluxes for galaxies means that radio surveys can serve as a useful
surrogate for such IR data. Below $10^{23}$ W Hz$^{-1}$, the radio-galaxy 
luminosity function is dominated by emission from galaxies driven by star 
formation. The VLA at 20cm has a field of view of 30 arcmin (FWHM) and can
detect radio sources down to about $10^{22}$ W Hz$^{-1}$ out to
$z=0.4$. Since the radio sky is much less crowded at these
levels than in a deep optical image, the optical identifications of galaxies
bright enough to be giant systems at the redshift of a cluster are very
likely to be associated with clusters at redshifts as large as 0.4. 

	In this paper we report optical observations of the bright radio
ID's found in two clusters, A2125 and A2645 imaged with the VLA at 20cm. The 
radio observations are reported by \cite{dw98}. These two
clusters were chosen as examples of rich systems studied by \cite{bo84}.
Both are richness class 4 clusters and both have redshifts
of $\sim 0.25$. However, A2125 has a large blue fraction ($0.19 \pm 0.03$)
while A2645 has a much smaller blue fraction ($0.03 \pm 0.05$). Does this
difference in blue fraction translate into a corresponding difference in
the radio galaxy population? Initial results for this project showed a large 
excess of radio galaxies in Abell 2125 compared with A2645 
(\cite{dw96,ow96,oetal96}). Now we consider the situation in more detail
using new observations. Below we report both optical imaging and spectroscopy of 
the radio identifications found in the directions of these these two clusters.

\section{Observations and Reductions}

	In August 1995 we imaged Abell 2125 and Abell 2645 with the 
KPNO 0.9m telescope in B and Cousins R. The CCD detector was a 2048X2048
Tektronics chip with a pixel size on the sky of 0.68 arcsec. Four twenty
minute, dithered exposures were stacked for both R images. In B one five 
minute exposure was obtained for A2645 and a ten-minute exposure for A2125. 
All frames were bias corrected and flattened using a twilight frame. Each frame 
was calibrated using about 20 standard stars (\cite{lan83,lan92}), 
bootstrapping 
from a shorter observation from a photometric night if the calibration for the
longer integration was questionable. Acceptable calibration nights had rms 
errors in the derived transformations of $<0.03$ magnitudes. The $B$-band
data fro A2125 were supplemented by a 20-minute exposure at the prime focus
of the KPNO 4m Mayall telescope using a similar filter/CCD combination,
calibrated using stars from the M92 field reported by \cite{c85}. Also
in November, 1998, Elizabeth Rizza obtained a deeper B frame of A2645
for us with one
hour total integration using the 0.9m with the same CCD and filter as 
our earlier observation. We have used our five minute, well-calibrated B 
frame to bootstrap the calibration to the deeper image and the colors 
we report for A2645 come from this image.

	The R observations transformed well without a color term to
Cousins R; however, the B observations from the 0.9m telescope needed 
an 8.6\% $B - R$ color
correction to match the Johnson B standard stars. This color term is not well
matched to redshifted galaxy energy distributions and thus we have
calculated the B magnitudes in the instrumental system by using the
Johnson B zero point for $ B - R = 0$. The instrumental filter + CCD
response has been calculated by \cite{m98} from the KPNO response
curve and this B system has been used in all model spectra calculations
we have performed. For the 4-m data, we have included correction for
a mild color term (0.04 magnitude in $B$ per unit of $B-R$).

	We have adopted the Gunn-Oke (\cite{go81}) metric aperture for measuring
galaxy magnitudes. For our adopted cosmology ($H_0 = 75$ km s$^{-1}$
and $q_0 = 0.1$), this corresponds to a radius of 13.1 kpc (3.8 arcsec for
$z=0.25$). This has the
advantage of containing almost all the light for galaxies with
luminosities $\le L_{*}$ (\cite{lo95}) while being small enough to 
minimize the error due the sky background and in most cases avoiding
contamination by nearby objects in the aperture.

	Magnitudes and internal errors were calculated using the IRAF
task PHOT. A calibration error of 0.05 magnitudes was adopted as an 
estimate of the total photometric calibration error, including the
transformation, any systematic error associated with the individual
observation and any systematic background errors. The errors in the
R magnitudes are all dominated by these external terms and an error
of 0.05 magnitudes is
adopted for all R observations. The color errors for faintest objects
are sometimes dominated by the counting statistics in B exposure. Thus 
for the colors we have quoted an error combining both the counting and the 
calibration errors.
	
	Spectra of the radio-galaxy identifications were obtained
in June 1995 and May 1996, using a long slit on the KPNO 4m Mayall telescope 
with the RC spectrograph. The grating had 316 lines/mm, blazed at 4000\AA\ . 
In 1995, the program was primarily intended for high-redshift radio galaxies 
and we covered the wavelength range 3500--7000\AA\  for this purpose. In 1996 
we chose a redder wavelength region of 4300--8600\AA\  to cover both the 
3727\AA\ [O II] line and the the [N II]/H$\alpha$ complex in the 
restframe of a $z=0.25$ cluster. For the 1996 observations,
we used the Risley prisms for atmospheric dispersion compensation, allowing
more freedom in selecting a position angle for the slit without compromising
signal in the blue. Slit widths 1.5-2" were used, for the best
contrast between galaxy and night-sky light at the relevant angular sizes.
When possible, the long slit was rotated so that two radio galaxies or a radio 
galaxy and an apparent companion could be covered in a single exposure of 5--15
minutes. Standard stars were observed each night to calibrate
the shape of the spectral energy distribution for each galaxy, although
no absolute calibration was attempted. 
Comparison of photometric and spectroscopically synthesized (B-R) values
shows that the flux calibration is consistent for continuum shape and
line ratios at better than the 10\% level from 4500-7000 \AA\ .

	Redshifts were obtained using the IRAF task, FXCOR, with a
K0 III radial velocity standard as the template. The calibration
errors were estimated using radial velocity standards observed on
different nights and was found to be $\le 20$ km s$^{-1}$. Thus the
errors are dominated by the individual spectral S/N. Without a
large database to further evaluate errors, we have adopted the
quoted errors from FXCOR.

\section{Results from Radio \& Optical Observations}
 
        The optical identification process is described in \cite{dw98}.
        Our spectroscopy is complete for radio identifications in these
clusters among galaxies brighter than Gunn-Oke $M_R =  -21.0$, 
except for four objects in A2645 and one in A2125 which would be 
brighter than this limit if they were at the cluster redshift. 
In table \ref{2125ids} and \ref{2645ids}, we summarize the 
magnitudes, colors, and heliocentric redshifts for each galaxy. An ``e''
appended to the redshift indicates the redshift is based on an
emission line spectrum. Errors are 
quoted in parenthesis. Some of our radio galaxies lie outside the area
covered by the CCD frames. Magnitudes for these objects, estimated from the
Digital Sky Survey with rough calibration form our CCD images, are given in 
parenthesis without any errors.

In table \ref{a2125p}, we present more detailed
information for each cluster member in A2125 and A2645 with spectroscopy:
absolute Gunn-Oke R magnitude, B-R color, D(4000), log of the absolute radio
luminosity, and for fairly isolated galaxies, a mean surface brightness and
concentration index (\cite{ab94}). 
In this table we have corrected the magnitudes and
colors for galactic extinction using the results of \cite{bh81} and applied
a K-correction (\cite{m98}). The surface brightness have been corrected for 
$(1+z)^4$ cosmological dimming as well.

In table \ref{a2125s}, we summarize the emission-line luminosities and 
calculated star formation rates for the A2125 members. In calculating the line
luminosities we have assumed that the line emission is actually spread 
throughout the entire galaxy rather than concentrated only in the nuclear 
region sampled by the slit.  To make this correction, we adjusted the absolute 
calibration of the spectra to match the Gunn-Oke aperture 
magnitudes calculated from the direct images.  The spectra were dereddened for 
galactic extinction, and K-corrections were estimated individually for each 
galaxy by comparing the flux over the B and R bandpasses for the galaxy in the
observed frame and artificially shifted to zero redshift. The K-corrections 
derived in this way are in reasonable agreement with \cite{fg94} over the 
range of galaxy morphologies in our sample based on the concentration 
parameters listed in table 3.

\section{Velocity Field and Mean Cluster Redshifts}

	Very few redshifts besides those presented here have been published
for these two clusters. For A2645, \cite{new88} measured five cluster
galaxies with adequate accuracy to use for an estimate of the
cluster mean velocity. One object is in common with our results and
agrees well within the estimated errors. In figures \ref{a2125v} and 
\ref{a2645v}, we show the velocity distributions near $z=0.25$ from our data, 
including \cite{new88} for A2645. For both clusters, there is a concentration 
near 0.25. For the combined sample for A2645, we find  a mean velocity of 
$0.2500\pm0.0020$. 
For A2125, we find a biweight scale (dispersion) of 891 km sec$^{-1}$ and a 
biweight mean of $0.2460\pm0.0007$. In A2125 there are three galaxies with 
redshifts near 0.275.  These galaxies do not survive a $3\sigma$ clipping with 
the calculated dispersion. Thus we find 27 radio identifications consistent 
with cluster membership in A2125. We adopt these mean redshifts for our further 
analysis of each cluster.

\section{Optical and X-ray Properties of A2125 and A2645}

	In figures \ref{a2125} and \ref{a2645}, we show the $R$ optical images 
of A2125  and A2645 with contours of the X-ray emission.
superimposed. The A2125 X-ray image is from the {\it ROSAT} PSPC while the
A2645 X-ray results are from {\it ROSAT} HRI.
The sensitivity of the HRI image of A2645, , smoothed to the PSPC resolution,
is significantly lower 
than A2125, but it is clear that the X-ray emission from A2645 is much more 
centrally condensed than A2125. This unusual property of A2125 is discussed in 
terms of a cluster-cluster merger by \cite{w98,r93}.

	In figures \ref{a2125} and \ref{a2645} we also include the positions of 
the radio galaxies reported here. One can see the difference in the apparent
density of radio objects and the correlation of the distribution in the
A2125 with the extended optical and X-ray structures. 

	Although these clusters were picked to have very
similar properties except for their blue fractions, a more detailed
look reveals significant differences. First, A2125 appears to be
a cluster merger in progress, as judged from the X-ray morphology (\cite{w98})
and the galaxy distribution. In contrast, A2645 appears to centrally condensed
and relatively relaxed. Its X-ray emission is also an order of magnitude
stronger and much simpler than A2125. Using the count rate in the inner
500 kpc and correcting statistically to the entire cluster (\cite{bh93}), we
find $L_x=3.6 \times 10^{44}$ ergs s$^{-1}$ for A2645 and $ 3.7 \times 10^{43}$ 
ergs s$^{-1}$ for A2125.  
In figures \ref{a2125a} and \ref{a2645a}, we show the
X-rays as a greyscale with the adaptively smoothed optical galaxy contours
overlayed (\cite{b91,m98}).
The adaptive kernel method is a two-step process. The initial
smoothing scale is set at 0.5 Mpc at the redshift of the cluster. The
final smoothing of the data is a function of the local galaxy surface
density derived from the initial smoothing. The estimated background
contribution to the galaxy surface density has been removed before
plotting the optical contours with the lowest contour set to the
estimated background level. One can see the excellent agreement between 
the X-rays and the optical galaxy counts.

	In order to estimate the richness of the cluster to compare with
other clusters, we counted galaxies down to a constant absolute
magnitude as evaluated in the cluster's emitted frame, and within
projected radii of 2 Mpc, $N_{2.0}$ (\cite{ab58,aco89}), and 0.5 Mpc, 
$N_{0.5}$ (Bahcall 1981), assuming $H_0 = 75$ km s$^{-1}$ Mpc$^{-1}$ and 
$q_0 = 0.1$.
The limiting absolute magnitude was $M_R = -20.5$. This
was determined by finding the absolute magnitude cutoff which agreed
best with the Abell counts for a set of lower redshift, very rich clusters
(\cite{m98}). Thus these results should be comparable with other
richness estimates for clusters while being based on more consistent 
and quantitative selection techniques than Abell's approach. 

	We find $N_{2.0} = 278$ and $232$ for A2125 and A2645,
in agreement with the estimates of \cite{ab58}. However, we find
$N_{0.5} = 42$ and $63$ respectively. If we define the compactness as
$C = N_{0.5}/N_{2.0}$, we find that A2645 is significantly more compact
with $C = 0.27$, while A2125 has $C = 0.15$. This difference in
concentration is significant at the 99\% level using a $2\times 2$
contingency test. Inside the 0.5 Mpc radius
we find a blue fraction for A2125 of $f_B = 0.20$ and for A2645 we find
$f_B = 0.05$,  in agreement with Butcher and Oemler (1984). See \cite{m98} for 
details. 
	 
	Thus, besides the difference in blue fraction, A2125 is much
less compact than A2645 in both the X-ray and the optical bands. Figures
\ref{a2125a} and \ref{a2645a} show this graphically. A2125 has two outlying 
clumps, consistent with the aftermath of a major merger while A2645 is much 
more compact and isolated as seen in both the X-ray gas and galaxy 
distributions.  

\section{Results for the Radio Galaxies}

	Several interesting differences between the populations in these
clusters can be seen from tables \ref{2125ids} and \ref{2645ids}. First, and 
most obvious, Abell 2125 has many more radio galaxies than Abell
2645. There are 27 radio galaxies associated with Abell 2125 compared with only 
four found in Abell 2645. Admittedly, our observations are more
incomplete for A2645, but this cannot account for the difference.
Within a circle 2.5 Mpc in diameter in projection at the cluster redshift,
we have redshifts for all ID's brighter than $m_R=19.5$ ($M_R = -21.0$), 
except for one object in A2125 and 5 in A2645. We find 25 confirmed cluster 
members for A2125 and only 2 for A2645. Thus if all the unmeasured objects in 
A2645 turn out to be cluster members, then we still find a large excess in 
A2125. From table \ref{a2125p}, all four confirmed A2645 radio galaxies have 
radio luminosities $\ge 10^{23}$ W Hz$^{-1}$ while only 4 of the A2125 
galaxies fall in this luminosity range. 
      For A2645, all the confirmed cluster members exhibit old stellar
populations, typical for their  radio luminosities.
Therefore the excess radio population in A2125 is 
entirely from lower-power objects, in the luminosity range 
$ < 10^{23}$ W Hz$^{-1}$. This might suggest star formation as the dominant 
source of the excess in A2125. However, their optical spectra present a more 
complex picture.

In A2125, there is a clean color demarcation between objects with emission lines
and pure absorption spectra, in the sense that all radio sources with observed
$(B-R)<1.6$ have emission lines and redder objects do not, suggesting that
most of these objects have line emission directly related to the observable
stellar populations and not, for example to AGN which are too weak to 
dominate the integrated colors. This also applies
to the foreground and background objects too, as it happens, though the
interpretation of the color is rather different at the lower foreground 
redshifts. There are no galaxies in the redshift-culled A2645 sample 
blue enough to have emission lines by this criterion, and indeed none
are observed to show emission lines. This demonstrates that the blue
radio-source population is uniquely associated with events in A2125;
the red population is numerically enhanced, while the blue one has
appeared from nowhere in comparison.

	In table \ref{a2125p}, we have divided the A2125 cluster radio galaxies
into four classes depending on their optical spectra. About half (13/27)
are dominated by old stellar populations based on their colors and/or
the 4000\AA\ break. None of these systems have detected line emission.
However, 8 of these galaxies have radio luminosities below $10^{23}$ W 
Hz$^{-1}$, a range normally associated with star-forming galaxies rather
than traditional AGN. A second group of 5 have spectra  consistent 
with vigorous star formation
based on colors, D(4000) and emission lines. One object has emission line 
ratios consistent with AGN activity
and a radio luminosity greater than is common for star forming objects. However,
its small value of D(4000) suggests star formation as well.

	The remaining 8 objects present the biggest puzzle. These objects
all have some evidence of a younger stellar population, either from their color,
low values of D(4000) or both. However, they have weak or 
undetectable line emission, which would in itself suggest weaker
star formation than would their radio luminosities. 
In figure \ref{ci}, we show the concentration index data from
table \ref{a2125p}. The plot shows that most of the old stellar
population objects cluster near the de Vaucouleurs profile while the
starforming objects are closer to the exponential disk curve. The
intermediate objects are spread over the entire range, although four
of the seven lie on the exponential disk curve. Thus most of the weak
radio galaxies have optical properties which are most consistent with
systems later than E/S0.

	In table \ref{a2125s} we consider the origin of the radio emission by
comparing the predicted star-formation rates (SFR) for the A2125 radio galaxies 
as based on H$\alpha$+[NII] luminosity, [O II] luminosity and radio luminosity. 
We use the values from \cite{ken98} (corrected for 1.1 magnitudes of extinction 
at H$\alpha$) for H$\alpha$ and [O II]. For the radio SFR we use the 
relationship in \cite{co92}. We plot the radio luminosity versus the line 
luminosities in figures \ref{ha} and \ref{oii}. The solid line indicates the 
prediction from \cite{ken98} and \cite{co92} for a constant SFR and the
dashed lines the expected uncertainty due to the local dust obscuration
variations. For 
H$\alpha$+[NII] and [OII] the obviously star-forming galaxies (with strong line
emission, from Table \ref{a2125p} agree with the predictions 
from the radio. For all the other object classes, the radio flux density is 
stronger than predicted from the line luminosities. 

	Also one can see from table \ref{a2125s} that for the old stellar 
population objects, the radio-derived SFR is much larger
than the limit we can set from the optical lines. This is consistent with
the radio emission being driven by radio jet emission related to weak AGN's.
For the stronger sources we can see extended radio jet structures which
confirm this picture. For the sources weaker than $10^{23}$ W Hz$^{-1}$
we see unresolved sources consistent with weak FR I sources but which
could, in principle, be very obscured star formation. But from the current
evidence, these objects seem most consistent with FR I/weak AGN activity.

	For the objects classified as starbursts, the agreement between the 
radio and optical SFR's is surprisingly good given the uncertainties in the 
calculations. Only 154040+661309 has some disagreement between the different
techniques in the sense that the radio is too weak. This might be explained
if the extinction is lower than assumed in this case and/or the star formation
is concentrated to the core of the galaxy, since our estimate of the absolute
luminosity assumes the EW is the same over the entire Gunn-Oke aperture.

	For the Intermediate sample the answer is less clear but in most
cases, the star formation rates do not disagree by factors larger than the
corrections applied. For the four systems whose light distributions are
consistent with disks, the disagreement with the  [O II] results are a factor
of 3-5. The radio SFR for 154053+660526 agrees well with the H$\alpha$
luminosity. The three other objects which do not fit a disk-model well have
no detected line emission, redder colors and larger D(4000)'s than the
other objects and thus seem to be candidates for weak AGN's. However, 
in all cases, it is hard to make a strong case for either mechanism. 
We could be looking at weak AGN's, very obscured star formation or a
mixture of the two. However, from the preponderance of the data we favor
star formation as the origin the radio emission for most of these systems.

	The AGN 154109+661544 stands out as a strong AGN. Alone among the
radio sources we have observed in A2125, it has a ratio of [NII]/H$\alpha$ line
strengths consistent with an AGN instead of star formation. Its radio
luminosity is clearly above the star-formation range and the radio
prediction of the the SFR is 50 times higher than the line luminosities
predict. However, its D(4000) and color are very small and consistent
with star formation as well as AGN activity or possibly with a very
strong non-thermal component.

\section{Discussion}

	The blue cluster Abell 2125  has a radio galaxy 
population below $10^{23}$ W Hz$^{-1}$ which is not found in the red
comparison cluster, Abell 2645. Part of this population shows active star
formation and lies preferentially on the outskirts of apparently merging clumps.
However, most of this excess appears to be associated with galaxies with
little or no evidence for star formation rates high enough to explain
the radio emission. Two logical possibilities exist; either the excess
radio emission is due to weak AGN activity, difficult to detect except
from its radio emission, or the star formation activity is well hidden,
probably by dust. 

	In either case the radio-galaxy population seems to be related to
the current nature of the clusters. A2645 is much more X-ray luminous and
seems to have a relatively simple structure resembling many cooling-flow 
clusters.
A2125 appears to be in the middle of a violent merger. Hydrodynamic
simulations have suggested that a cluster merger can disrupt a cooling
flow and heat the X-ray gas, so that unusually weak X-ray emission could
be a side effect of such an event ({\it e.g.} \cite{g96,r98}), although
it is not clear whether this process is just beginning or is in its
later stages (see e.g. \cite{r93}). This raises the question
of whether the merger has anything to do directly with the radio excess. One
might consider accretion of radio galaxies from the field, whose activity 
have not yet been snuffed out by the intracluster environment, but 
it is difficult 
to understand why we do not see such a population in the outskirts of A2645. 

	Another appealing possibility is that the cluster-cluster interaction
has somehow enhanced the radio population, either through mutual
galaxy interactions or through galaxy-ICM interaction. The question is
how this might be happening in detail. If galaxy-galaxy mergers/interactions
are responsible, where and how are the interactions taking place ?
Hyperbolic encounters one might expect in cluster-cluster merger are
more likely to inhibit star formation, if local examples may be generalized
(\cite{c88}). From figure \ref{a2125} one can see that the starforming
and intermediate radio galaxies mostly avoid the central X-ray/optical
concentration. However, they follow the general outline of the
of the extended cluster to the southwest, apparently lying on the outskirts
of the overall distribution. Thus it seems likely that whatever mechanism is 
responsible for the radio emission takes place outside the dense cluster core. 
From the Las Campanas optical spectral survey, \cite{ha98} find
a higher probability of line emission from galaxies in intermediate
density regions than in the dense cores of clusters or in the low density
field. This could be due to the cross-section for galaxy-galaxy interactions
being maximized in such an environment. \cite{l92} have pointed out the
existence of a surprisingly large population of close, blue pairs 
among the blue populations in Butcher-Oemler clusters which may
indicate some process for increased interactions between galaxies outside
the cluster core. To 
increase galaxy-galaxy interactions, one needs regions with low relative
velocities between galaxies and as high galaxy densities as possible.
Such regions might occur in the outlying groups of each of
the two subclumps which must also be undergoing group-group mergers. Whether
these occur at low enough relative velocities is an open question. Another
possibility is that a cluster-cluster merger simply correlates with a
denser supercluster region which would maximize the galaxy-galaxy
interaction rate. These issues probably require studies of simulations
of cluster-cluster mergers to understand if any such process is taking
place.

	However, the lack of the star formation signature in the optical
spectra of many of the radio galaxies might mean that AGN's as well as
starbursts have been stimulated. The origin of AGN's is still unclear but
it could be that AGN activity is another result of galaxy-galaxy interactions
or mergers, perhaps with a different timescale ({\it e.g.}  \cite{wil96}). 
Many more clusters will need to be studied, and more detailed work on
these clusters needs to be done, to explore the correlations and
the physical picture suggested here. As a next step, we will present
a more detailed study of the properties of the radio-loud and radio-quiet 
galaxies in A2125 in a subsequent paper.

\section{Conclusions}

\begin{enumerate}
	\item The blue Butcher-Oemler cluster, A2125, has a large excess of
radio galaxies relative to the red cluster, A2645, although both clusters
have similar richnesses and redshifts. 

	\item This excess occurs for 20cm radio luminosities below
$10^{23}$ W Hz$^{-1}$. Locally, radio galaxies at these luminosities
are mostly driven by star-formation.

	\item The optical properties of these radio galaxies are mostly
consistent with systems later in type than E/S0. However, most have lower
line luminosities than would be expected from their radio luminosity if
star-formation is driving the activity. This result seems most consistent
with enhanced extinction relative to most nearby systems obscuring the
star-forming regions. 

	\item Besides A2125 blue galaxy population, the cluster appears
to be in the process of a cluster-cluster merger which may be related to
the unusual activity.
\end{enumerate}

\acknowledgments

	The authors wish to thank Elizabeth Rizza for obtaining the
B image of A2645 for us with the KPNO 0.9m.  

\clearpage
\begin{deluxetable} {rrrrrrrr}
\tablecaption{Optical IDs for Abell 2125 \label{2125ids}}
\startdata
&\colhead{R.A.(2000.0)}&\colhead{Dec(2000.0)}&\colhead{R(GO)}&\colhead{B$-$R}&
\colhead{Redshift}&\nl
&15 39 08.82&66 08 54.4&(18.8)&\omit&0.2448(0.0003)&\nl
&15 39 33.07&66 07 43.5&20.00&2.11(0.09)&\nl
&15 39 46.83&66 06 24.9&20.92&\nl
&15 39 59.33&66 11 26.7&18.71&1.43(0.05)&0.2457e(0.0003)&\nl 
&15 39 59.40&66 16 07.4&18.02&2.34(0.05)&0.2459(0.0003)&\nl 
&15 40 00.00&66 05 51.6&20.38&0.98(0.11)&\nl
&15 40 05.35&66 10 13.0&18.03&2.37(0.05)&0.2425(0.0003)&\nl
&15 40 08.70&66 15 36.3&18.99&0.74(0.08)&0.0931(0.0003)&\nl
&15 40 09.05&66 12 17.0&17.63&2.41(0.05)&0.2455(0.0003)&\nl
&15 40 12.06&66 12 09.9&18.11&2.33(0.05)&0.2567(0.0003)&\nl
&15 40 15.85&66 11 09.9&18.62&1.31(0.08)&0.2449(0.0003)&\nl
&15 40 26.00&66 30 30.4&(18.5)&\omit&0.2573(0.0004)&\nl
&15 40 30.13&66 12 14.3&19.08&1.79(0.05)&0.2458(0.0004)&\nl
&15 40 30.23&66 13 06.3&18.15&2.23(0.05)&0.2496(0.0004)&\nl
&15 40 30.91&66 12 26.3&19.50&1.35(0.06)&0.2458e(0.0003)&\nl
&15 40 33.62&66 08 01.5&20.90&0.93(0.08)&\nl
&15 40 40.03&66 13 09.0&18.87&1.19(0.05)&0.2462e(0.0003)&\nl
&15 40 40.83&66 26 54.2&16.46&1.56(0.07)&0.0564(0.0003)&\nl
&15 40 42.55&66 08 54.2&20.48&1.82(0.28)&\nl
&15 40 43.18&66 10 20.8&20.08&1.77(0.16)&\nl
&15 40 49.18&66 18 39.7&18.52&2.21(0.05)&0.2433(0.0003)&\nl
&15 40 50.98&66 16 32.2&18.34&1.56(0.05)&0.1277(0.0002)&\nl
&15 40 51.80&66 06 31.0&18.43&2.55(0.06)&0.3640(0.0003)&\nl
&15 40 53.64&66 05 26.4&19.43&1.65(0.10)&0.2527(0.0004)&\nl
&15 40 54.51&66 11 27.5&19.53&\nl
&15 40 54.66&66 17 15.7&18.43&1.89(0.05)&0.2459(0.0003)&\nl
&15 40 56.88&66 26 45.8&18.25&1.70(0.08)&0.2443(0.0003)&\nl
&15 41 00.52&66 13 54.2&20.42&1.43(0.16)&\nl
&15 41 01.93&66 16 26.6&17.50&2.34(0.05)&0.2456(0.0003)&\nl
&15 41 05.33&66 12 35.7&18.89&1.54(0.05)&0.2754e(0.0003)&\nl
&15 41 07.29&66 21 23.9&20.81&0.75(0.15)&\nl
&15 41 09.73&66 15 44.5&18.75&1.23(0.05)&0.2526(0.0003)&\nl
&15 41 10.63&66 13 13.9&19.93&1.71(0.14)&\nl
&15 41 14.38&66 15 56.8&17.26&2.37(0.05)&0.2520(0.0003)&\nl
&15 41 14.60&66 21 39.9&16.58&1.27(0.07)&0.0855(0.0003)&\nl
&15 41 14.85&66 16 03.6&17.29&2.37(0.05)&0.2470(0.0002)&\nl
&15 41 15.36&66 15 58.9&17.80&2.41(0.08)&0.2466(0.0002)&\nl
&15 41 28.12&66 13 23.6&17.34&0.81(0.05)&0.0931(0.0003)&\nl
&15 41 33.91&66 22 54.7&19.55&2.31(0.16)&\nl
&15 41 33.98&66 31 11.3&(16.9)&\omit&0.2361(0.0004)&\nl
&15 41 40.99&66 22 37.8&20.41&0.53(0.10)&\nl
&15 41 43.22&66 15 16.6&18.56&2.22(0.08)&0.2520(0.0003)&\nl
&15 41 48.79&66 11 37.0&19.82&$>$2.0&\nl
&15 42 02.76&66 15 55.1&19.31&2.15(0.13)&0.2407(0.0003)&\nl
&15 42 03.88&66 26 31.7&19.14&2.13(0.11)&0.2464(0.0004)&\nl
&15 42 14.67&66 30 04.0&(18.6)&\omit&0.2999(0.0004)&\nl
&15 42 24.26&66 19 58.4&18.02&2.30(0.08)&0.2473(0.0003)&\nl
&15 42 24.67&66 29 04.3&(19.3)&\omit&0.2999(0.0004)&\nl
&15 42 25.06&66 09 17.3&20.94&1.16(0.21)&\nl
&15 42 41.09&66 24 36.9&18.08&0.97(0.07)&0.2757(0.0002)&\nl
&15 42 41.44&66 24 33.7&18.37&0.81(0.07)&0.2748(0.0002)&\nl
&15 43 01.07&66 15 16.9&19.92&\nl
&15 43 07.81&66 13 43.9&(19.3)&\nl
&15 43 44.83&66 30 01.0&(18.9)&\omit&0.2482(0.0004)&\nl
\enddata
\end{deluxetable}
\begin{deluxetable} {rrrrrrrr}
\tablecaption{Optical IDs for A2645 \label{2645ids}}
\startdata
&\colhead{R.A.(2000.0)}&\colhead{Dec(2000.0)}&\colhead{R(GO)}&\colhead{B$-$R}&
\colhead{Redshift}&\nl
&23 40 27.82&-09 09 45.9&(15.6)&\omit&0.0728e(0.0001)&\nl
&23 40 39.56&-08 51 37.1&20.11&2.19(0.09)&\omit&\nl
&23 40 43.18&-08 53 56.6&19.23&2.61(0.07)&\omit&\nl
&23 40 45.51&-09 01 55.0&19.61&2.47(0.08)&\omit&\nl
&23 40 49.39&-08 57 03.7&18.32&1.68(0.05)&\omit&\nl
&23 40 53.08&-09 03 15.3&18.86&0.62(0.05)&1.182e(0.0004)&\nl
&23 40 53.22&-08 58 30.1&18.88&1.62(0.06)&0.2945e(0.0002)&\nl
&23 40 57.04&-09 01 53.7&17.77&1.65(0.05)&0.1522(0.0003)&\nl
&23 41 07.73&-08 55 45.8&19.99&1.61(0.07)&\omit&\nl
&23 41 10.19&-08 57 11.9&18.03&2.39(0.06)&0.2477(0.0003)&\nl
&23 41 11.40&-08 44 47.5&(17.7)&\omit&0.2420(0.0004)&\nl
&23 41 11.57&-09 02 02.2&18.39&2.48(0.06)&0.2518(0.0003)&\nl
&23 41 13.04&-09 02 12.4&20.64&2.30(0.14)&\omit&\nl
&23 41 13.99&-09 07 47.8&18.93&2.50(0.06)&0.4422(0.0006)&\nl
&23 41 15.28&-09 05 17.2&18.04&2.53(0.05)&\nl
&23 41 16.20&-08 55 09.6&19.64&1.60(0.06)&\nl
&23 41 17.11&-09 02 17.0&19.55&1.63(0.06)&\nl
&23 41 20.76&-08 55 42.1&18.78&1.63(0.05))&0.1724e(0.0002)&\nl
&23 41 21.97&-09 02 13.5&18.63&1.31(0.05)&0.1071e(0.0002)&\nl
&23 41 26.03&-09 04 50.3&17.44&1.90(0.05)&0.0713(0.0006)&\nl
&23 41 37.33&-08 51 53.4&19.80&2.11(0.08)&\omit&\nl
&23 41 37.54&-09 11 00.0&20.09&$>3.6$&\omit&\nl
&23 41 37.86&-09 04 11.7&20.38&$>3.3$&\nl
&23 41 39.45&-09 02 11.4&17.40&1.90(0.05)&0.1539(0.0005)&\nl
&23 41 39.61&-08 48 36.8&(19.0)&\omit&\omit&\nl
&23 41 40.65&-08 43 13.2&(16.1)&\omit&0.0739e(0.0001)&\nl
&23 41 42.30&-09 14 32.3&(16.7)&\omit&0.0887(0.0002)&\nl
&23 41 44.10&-09 08 00.4&19.80&2.70(.11)&\omit&\nl
&23 41 57.99&-08 55 16.2&17.19&2.33(0.05)&0.1742(0.0006)&\nl
&23 41 59.74&-09 02 05.8&19.47&2.09(0.07)&\omit&\nl
&23 41 59.76&-09 03 48.5&16.91&2.17(0.05)&0.1529(0.0003)&\nl
&23 42 00.31&-09 03 00.4&20.18&2.13(0.09)&\omit&\nl
&23 42 03.04&-09 04 25.5&17.68&1.90(0.05)&0.1741(0.0003)&\nl
&23 42 03.30&-08 51 28.0&(19.1)&\omit&0.2533(0.0003)&\nl
&23 42 04.94&-09 15 10.9&(17.8)&\omit&0.2145(0.00004)&\nl
\enddata
\end{deluxetable}
\begin{deluxetable} {rrrrrrrrrr}
\tablecaption{A2125 \& A2645 Radio Galaxy Parameters \label{a2125p}}
\startdata
&\colhead{Source}&\colhead{$M_R$}&\colhead{B$-$R}&\colhead{D(4000)}&
\colhead{$log L_{20}$}&\colhead{C.I.}&\colhead{$\mu$}&\nl
&\omit&\omit&\omit&\omit&ergs s$^{-1}$&\omit&\omit&\nl
&\nl
&\colhead{Old}&\nl
&153908+660854&(-21.7)&\omit&2.13&24.96&\nl
&153959+661607&-22.48&2.42&1.98&24.13&0.53&22.74\nl
&154005+661013&-22.47&2.40&2.10&22.85&0.63&22.38\nl
&154009+661217&-22.87&2.48&1.96&22.79&0.62&22.89\nl
&154012+661209&-22.39&2.41&2.09&22.48&0.68&22.32\nl
&154026+663030&(-22.0)&\omit&2.18&22.58&\nl
&154049+661839&-21.98&2.44&2.05&24.36&\nl
&154101+661626&-23.00&2.46&2.12&22.47&\nl
&154114+661556&-23.24&2.51&2.12&23.64&\nl
&154114+661603&-23.21&2.54&2.21&22.87&\nl
&154115+661558&-22.70&2.46&2.18&24.35&\nl
&154133+663111&(-23.6)&\omit&2.03&22.89&\nl
&154224+661958&-22.48&2.35&2.08&22.45&0.61&22.29&\nl
&\nl
&\colhead{Starburst}&\nl
&153959+661126&-21.79&1.44&1.22&22.58&0.38&22.08&\nl
&154015+661109&-21.88&1.42&1.27&22.66&0.48&22.27&\nl
&154030+661214&-21.42&1.92&1.45&22.42&0.50&22.22&\nl
&154030+661230&-21.89&1.77&1.51&22.33&\nl
&154040+661309&-21.63&1.26&1.22&22.30&0.37&22.19&\nl
&\nl
&\colhead{Intermediate}&\nl
&154030+661306&-22.35&2.24&1.93&22.32&0.62&22.51&\nl
&154053+660526&-21.07&1.70&1.68&22.45&0.33&22.48&\nl
&154054+661715&-22.07&1.89&1.99&22.63&0.49&22.51&\nl
&154056+662645&-22.05&1.75&1.77&22.57&0.38&22.63&\nl
&154143+661516&-21.94&2.27&1.86&22.38&0.58&22.54&\nl
&154202+661534&-21.19&2.20&1.62&22.29&0.37&22.47&\nl
&154203+662631&-21.36&2.17&1.43&22.66&0.34&22.40&\nl
&154344+663001&(-21.6)&\omit&1.63&22.90&\nl
&\nl
&\colhead{AGN}&\nl
&154109+661544&-21.75&1.15&1.23&24.53&0.42&22.18&\nl
&\nl
&\colhead{A2645}&\nl
&234110--085711&-22.48&2.67&1.95&23.00&\nl
&234111--084447&(-22.8)&\omit&1.96&24.02&\nl
&234111--090202&-22.12&3.19&2.27&23.59&\nl
&234203--085128&(-21.4)&\omit&2.11&23.27&\nl
\enddata
\end{deluxetable}
\clearpage
\begin{deluxetable} {rrrrrrrrrr}
\tablecaption{A2125 Star Formation Estimates \label{a2125s}}
\startdata
&\colhead{Source}&\colhead{$log L_{H\alpha}$}&\colhead{SFR}&
\colhead{$log L_{O[II]}$}&\colhead{SFR}&\colhead{$log L_{20}$}&
\colhead{SFR}&\nl
&\omit&ergs s$^{-1}$&$M_{\odot}$ yr$^{-1}$&ergs s$^{-1}$&$M_{\odot}$ yr$^{-1}$
&W Hz$^{-1}$&$M_{\odot}$ yr$^{-1}$&\nl
&\nl
&\colhead{Old}&\nl
&153908+660854&\omit&\omit&$<40.35$&$<0.9$&24.91&2000&\nl
&153959+661607&$<41.02$&$<2.3$&$<40.49$&$<1.2$&24.08&300&\nl
&154005+661013&\omit&\omit&$<41.02$&$<2.3$&22.80&16&\nl
&154009+661217&$<41.00$&$<2.2$&$<40.47$&$<1.2$&22.75&14&\nl
&154012+661209&$<40.68$&$<1.1$&$<40.30$&$<0.8$&22.44&7&\nl
&154026+663030&$<41.07$&$<2.6$&$<40.74$&$<2.2$&22.54&9&\nl
&154049+661839&$<40.85$&$<1.6$&$<40.43$&$<1.0$&24.31&500&\nl
&154101+661626&\omit&\omit&$<40.24$&$<0.7$&22.42&7&\nl
&154114+661556&\omit&\omit&$<40.54$&$<1.4$&23.60&100&\nl
&154114+661603&$<41.21$&$<3.6$&$<40.81$&$<2.6$&22.83&17&\nl
&154115+661558&\omit&\omit&$<40.48$&$<1.2$&24.30&500&\nl
&154133+663111&\omit&\omit&$<40.87$&$<2.9$&22.84&17&\nl
&154224+661958&\omit&\omit&$<40.58$&$<1.5$&22.41&6&\nl
&\nl
&\colhead{Starburst}&\nl
&153959+661126&41.76&13&41.42&11&22.54&9&\nl
&154015+661109&41.81&14&41.34&9&22.59&10&\nl
&154030+661214&41.50&7&41.07&5&22.40&6&\nl
&154030+661230&41.24&4&41.45&9&22.36&6&\nl
&154040+661309&41.80&14&41.76&23&22.25&4&\nl
&\nl
&\colhead{Intermediate}&\nl
&154030+661306&$<41.04$&$<2.4$&$<40.46$&$<1.1$&22.28&5&\nl
&154053+660526&41.25&4&40.66&1.8&22.41&6&\nl
&154054+661715&\omit&\omit&$<40.51$&$<1.3$&22.58&10&\nl
&154056+662645&$<40.96$&$<2.0$&$<40.61$&$<1.6$&22.53&8&\nl
&154143+661516&$<41.17$&$<3.3$&$<40.51$&$<1.3$&22.33&5&\nl
&154202+661534&\omit&\omit&40.39&1.0&22.24&4&\nl
&154203+662631&\omit&\omit&40.82&2.6&22.61&10&\nl
&154344+663001&\omit&\omit&40.49&1.2&22.85&18&\nl
&\nl
&\colhead{AGN}&\nl
&154109+661544&41.76&14&41.54&14&24.48&750&\nl
\enddata
\end{deluxetable}
\clearpage
\centerline{\bf FIGURE CAPTIONS}
\figcaption[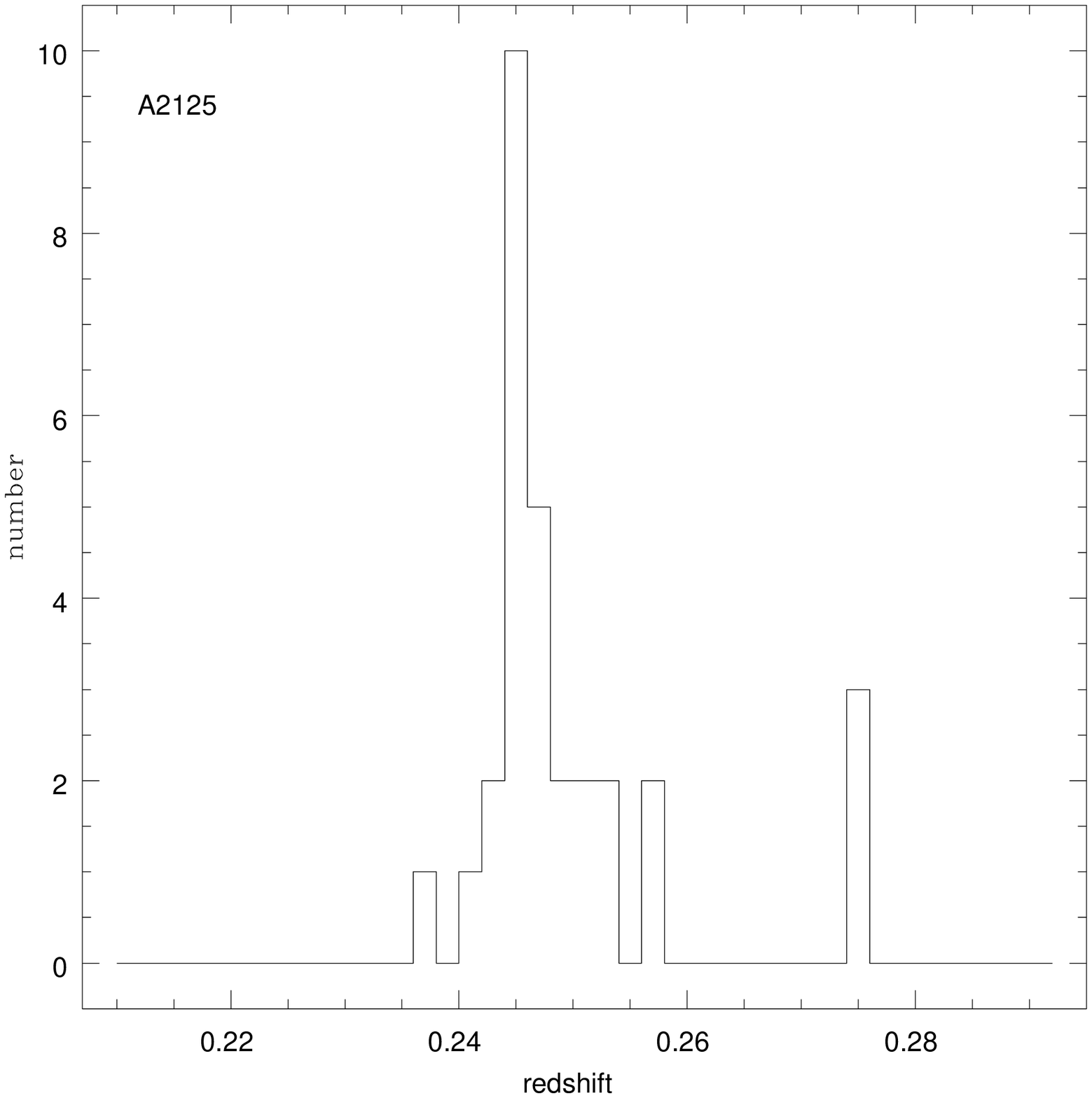]{Measured Redshift Distribution for A2125 
\label{a2125v}}
\figcaption[owenrev.fig2..ps]{Measured Redshift Distribution for A2645
\label{a2645v}}
\figcaption[owenrev.fig3.ps]{R Image of A2125 with contours of X-ray intensity
from ROSAT PSPC image overlayed. The X-ray image has been smoothed
to a resolution of 250 kpc at the redshift of the cluster. The radio
ID's have been marked on the image according to our spectral
classification: Old stellar populations (circles), starbursts
(7-pointed stars), intermediate (squares), and AGN's (diamonds).
Two of the radio ID's in Table \ref{a2125p} are outside the image field-of-view
(4.9 Mpc). Note that the central galaxy, near 154115+6616, is a triple system
and each of the three optical nuclei is a separate radio source 
(see Dwarakanath \& Owen 1999). \label{a2125}}
\figcaption[owenrev.fig4.ps]{R Image of A2645 with contours of X-ray intensity
from ROSAT HRI image overlayed. The HRI image has been regridded to
match that of the PSPC, and smoothed to a resolution of 250 kpc.
Three of the four radio ID's are marked on the image; the other is
outside the field-of-view. All three radio ID's are old-population based
on their spectra.
\label{a2645}}
\figcaption[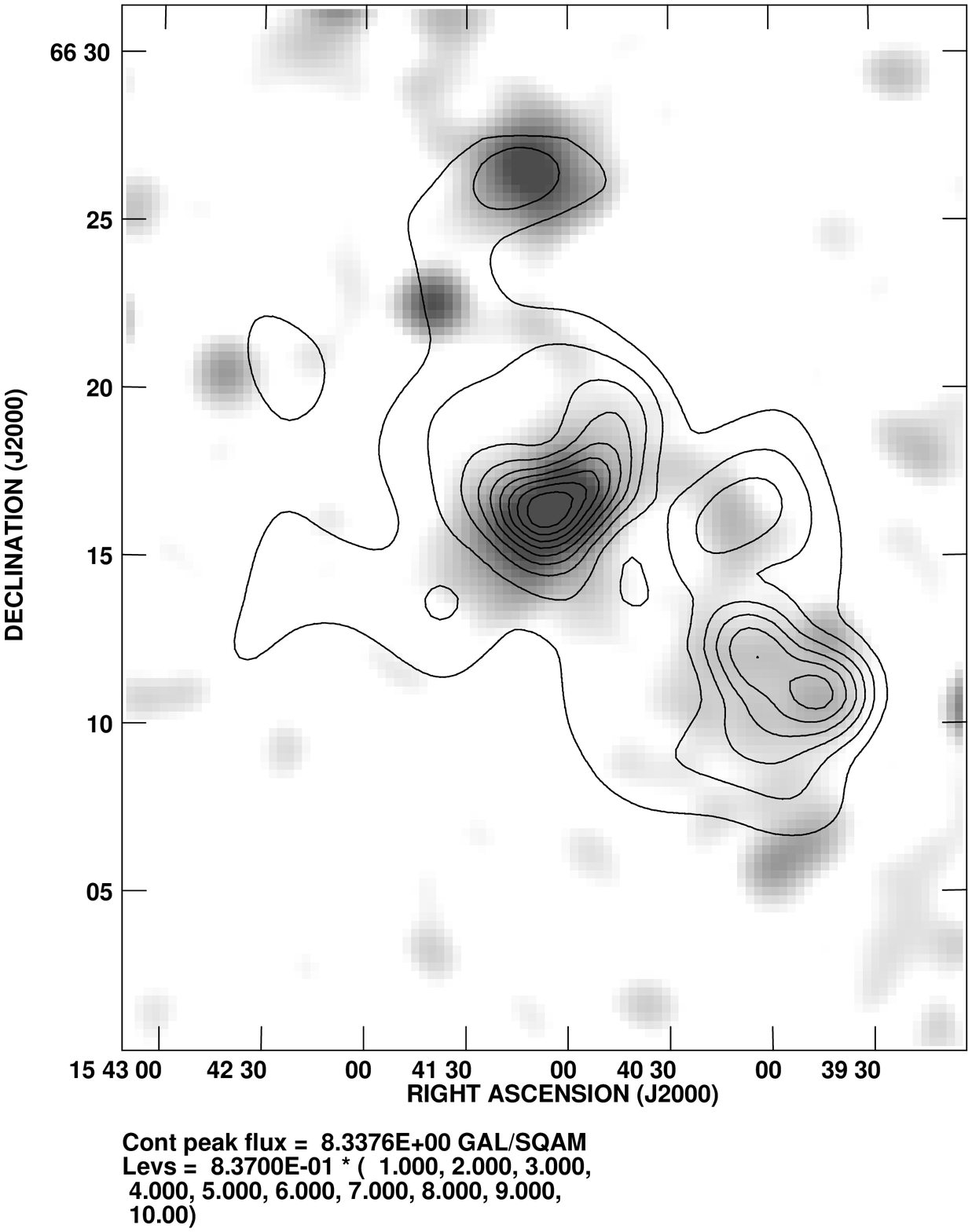]{Adaptively Smoothed Contours of Galaxy Density
overlayed on Greyscale of X-ray Intensity for A2125. \label{a2125a}}
\figcaption[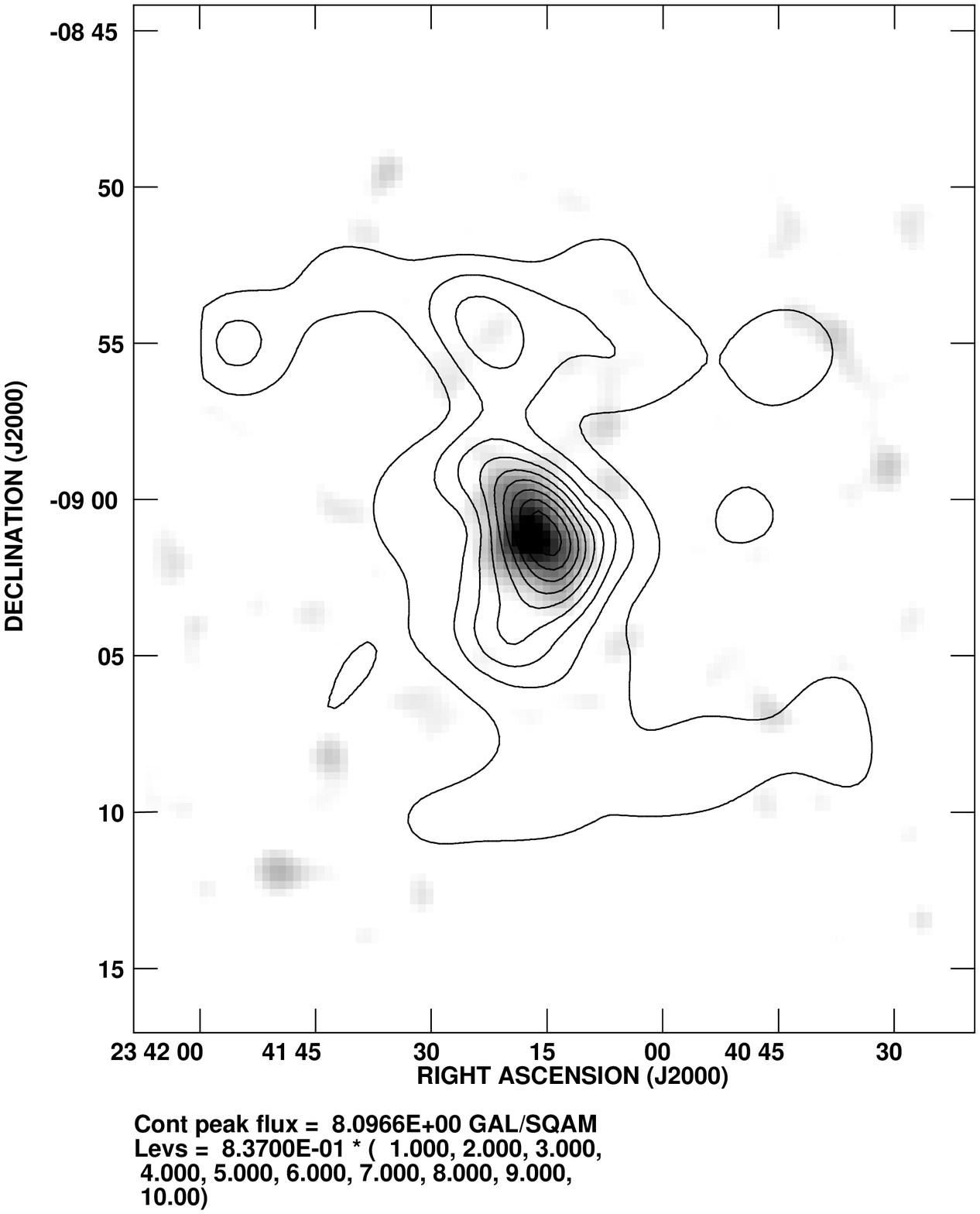]{Adaptively Smoothed Contours of Galaxy Density
overlayed on Greyscale of X-ray Intensity for A2645. \label{a2645a}}
\figcaption[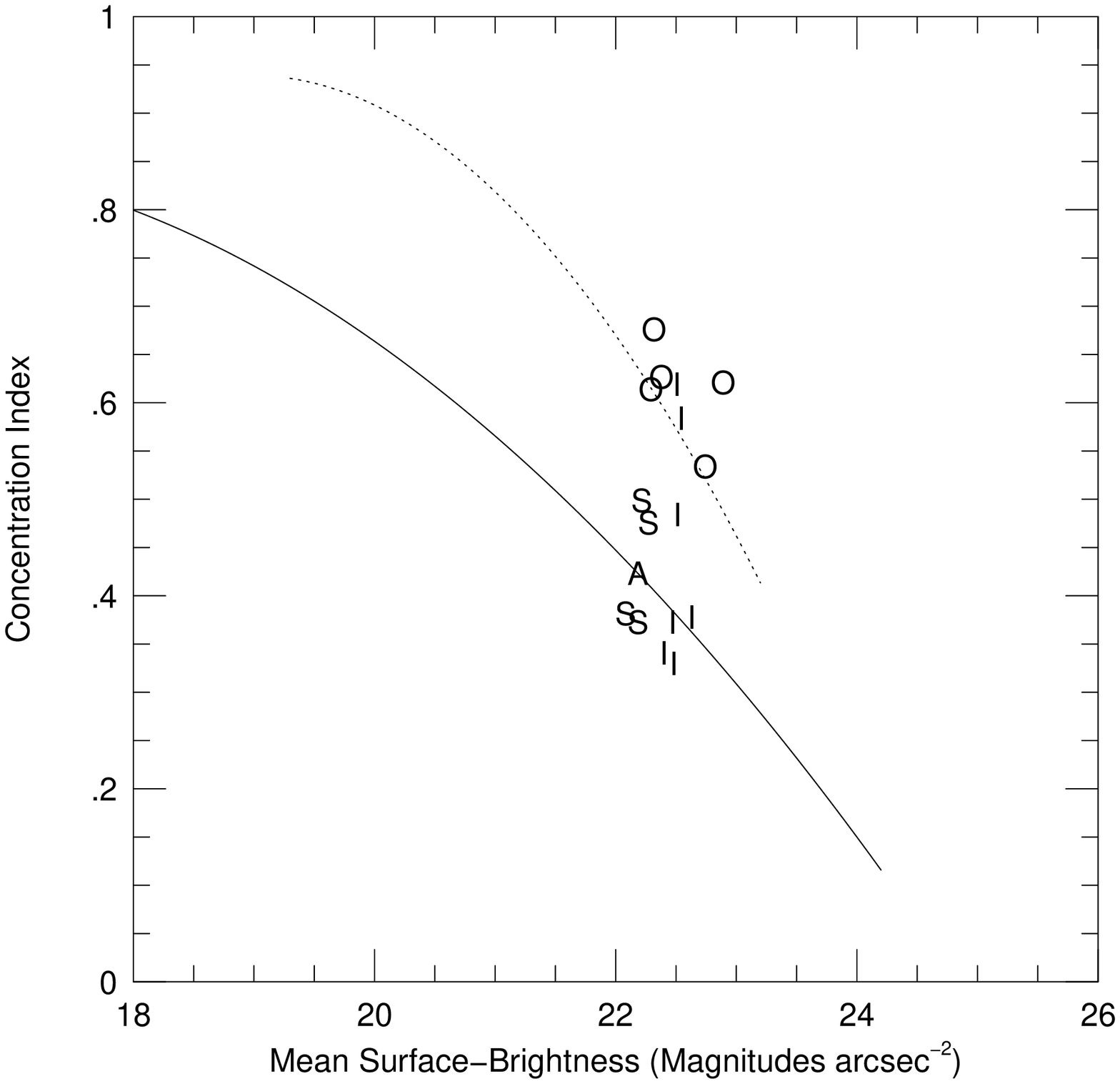]{Surface brightness versus Concentration Index for fairly
isolated galaxies. The symbols are as in figure \ref{ha}. The solid line
is the result for an exponential disk and the dotted line is for a de
Vaucouleurs' $r^{1/4}$ law. \label{ci}}
\figcaption[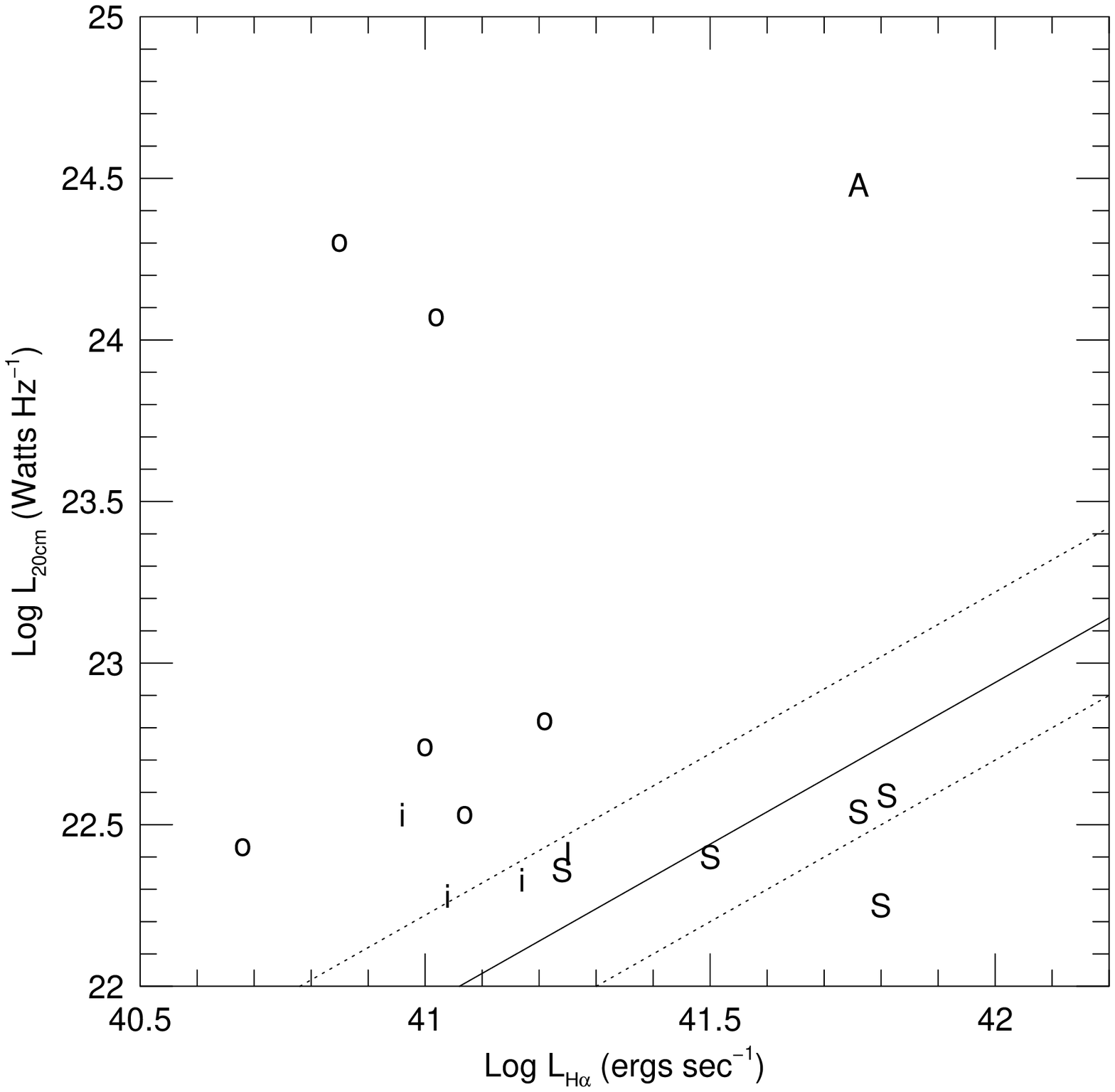]{20cm radio luminosity versus H$\alpha$+[NII] 
luminosity for the A2125 radio galaxies in Table \ref{a2125p}. The symbols 
represent the different classes: Old Population (O),
Starbursts (S), AGN (A), and Intermediate (I). Lower-case letters indicate
upper limits. The solid-line shows the result of
combining the Condon (1992) and Kennicutt (1998) relations for the SFR
assuming 1.1 magnitudes of dust extinction. The dashed-lines indicate the 
range of possible SFR's due to the uncertainty in the dust extinction
(0.5 to 1.8 magnitudes). 
\label{ha}}
\figcaption[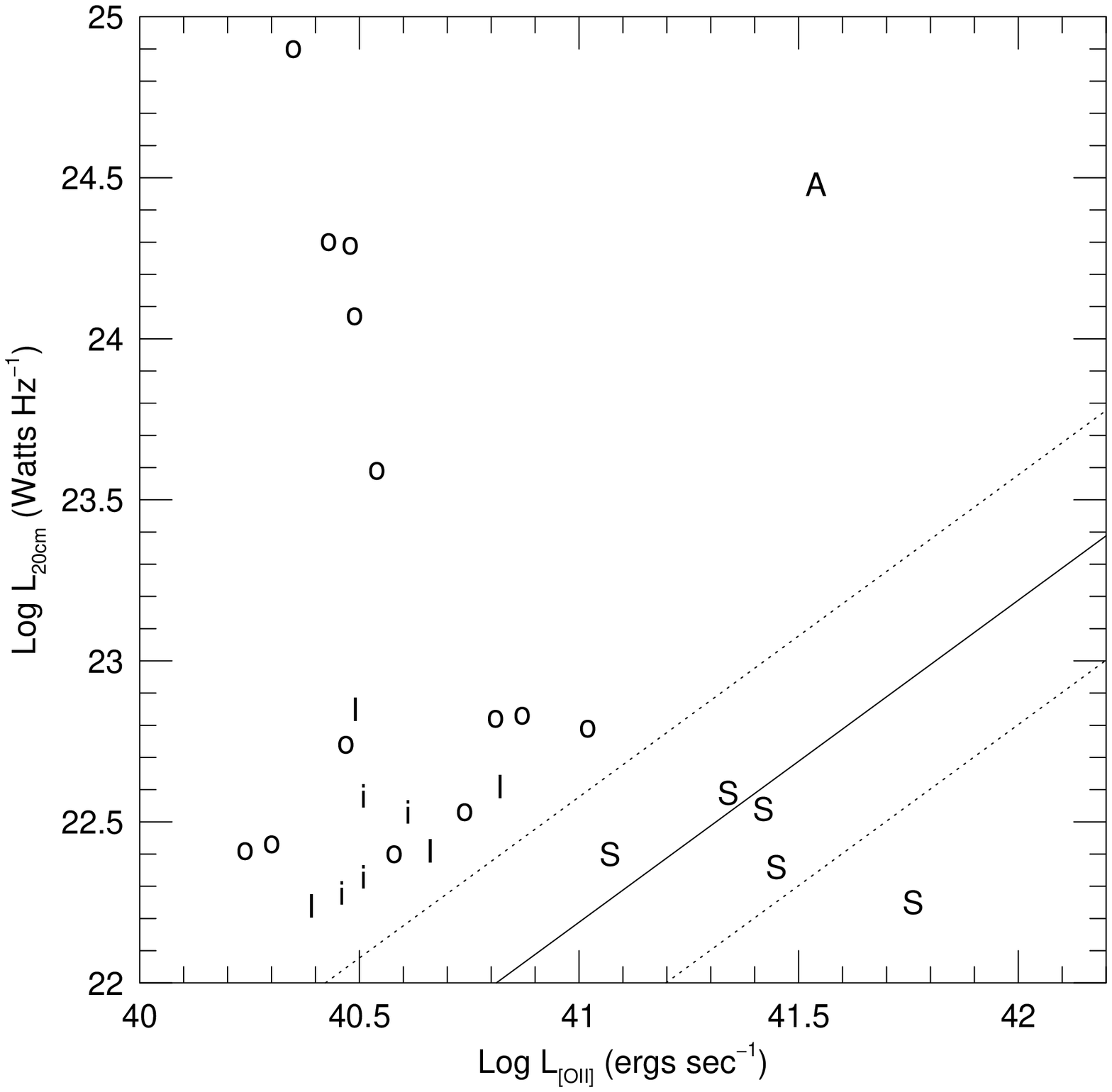]{20cm radio luminosity versus [OII] luminosity. 
Labels and the line are as referenced in \ref{ha}. The dashed-lines indicate
both the uncertainty in the dust extinction (0.5-1.8 magnitudes) as well as
an intrinsic scatter (28\%) in the [OII] SFR estimates (Kennicutt 1998).
\label{oii}}
\clearpage
\begin{figure}
\epsscale{2.0}
\plotone{owenrev.fig1.ps}
\end{figure}
\clearpage
\begin{figure}
\epsscale{2.0}
\plotone{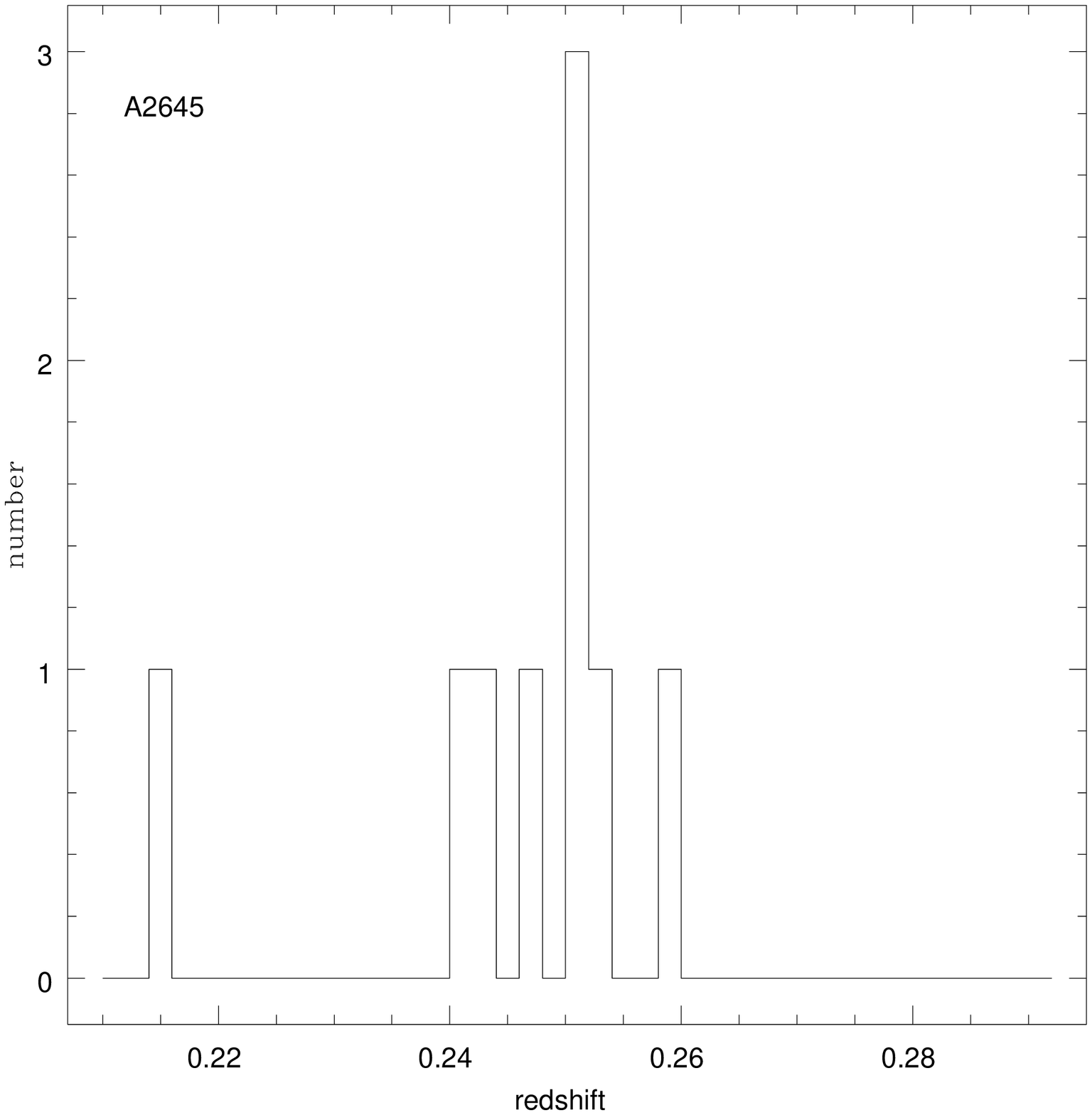}
\end{figure}
\clearpage
\begin{figure}
\epsscale{2.0}
\plotone{owenrev.fig5.ps}
\end{figure}
\clearpage
\begin{figure}
\epsscale{2.0}
\plotone{owenrev.fig6.ps}
\end{figure}
\clearpage
\begin{figure}
\epsscale{2.0}
\plotone{owenrev.fig7.ps}
\end{figure}
\clearpage
\begin{figure}
\epsscale{2.0}
\plotone{owenrev.fig8.ps}
\end{figure}
\clearpage
\begin{figure}
\epsscale{2.0}
\plotone{owenrev.fig9.ps}
\end{figure}
\clearpage

\clearpage	
\begin{thebibliography}{}
\bibitem[Abell 1958]{ab58} Abell, G. O. 1958, \apjs, 3, 211
%
\bibitem[Abell, Corwin \& Olowin 1989]{aco89}
Abell, G. O., Corwin, H. G., \& Olowin, R. P. 1989, \apjs, 70, 1
%
\bibitem[Abraham et al 1994]{ab94} Abraham, R. G., Valdes, F., Yee, H. K. C.,
\& van den Bergh 1994, \apj, 432, 75
%
\bibitem[Bahcall 1981]{Ba81} Bahcall, N. A. 1981, \apj, 247, 787
%
\bibitem[Beers et al]{b91} Beers,T. C., Forman, W., Huchra, J P.,
Jones C., \&  Gebhardt, K. 1991, \aj, 102, 1581
%
\bibitem[Briel \& Henry 1993]{bh93} Briel \& Henry 1993, \aap, 278, 79
%
\bibitem[Burstein \& Heiles 1981]{bh81}
Burstein, D. \& Heiles, C. 1981, \apj, 87, 1165
%
\bibitem[Butcher \& Oemler (1984)]{bo84}
Butcher, H. \& Oemler, A. 1984, \apj, 285, 426
%
\bibitem[Christian et al (1985)]{c85}
Christian, C.A., Adams, M., Barnes, J.V., Hayes, D.S., Butcher, H.,
Mould,  J. R., \& Siegel, M. 1985, \pasp, 97, 363
%
\bibitem[Combes et al (1988)]{c88}
Combes, F., Dupraz, C., Casoli, F., \& Pagani, L. 1988, \aap, 203, L9
%
\bibitem[Condon (1992)]{co92} Condon, J. J. 1992, \araa, 30, 575
%
\bibitem[Dwarakanath \& Owen 1996]{dw96}
Dwarakanath, K. S. \& Owen, F. N. 1996, in {\it Cold
Gas at High Redshift}, (Bremer et al, eds.), Kluwer, p183
%
\bibitem[Dwarakanath \& Owen (1999)]{dw98}
Dwarakanath, K. S. \& Owen, F. N. (1999), submitted
%
\bibitem[Frei \& Gunn (1994)]{fg94}
Frei, Z, \& Gunn, J. E. 1994, \aj, 108, 1476
%
\bibitem[Gomez et al 1996]{g96}
Gomez, P. L., Loken, C., Burns, J. O., \& Roettiger, K. 1996, \baas, 189, 11503
%
\bibitem[Gunn \& Oke 1981]{go81}
Gunn, J.E. \& Oke, J. B. 1975, \apj, 195, 255
%
\bibitem[Hashimoto et al.,\ (1998)]{ha98}
Hashimoto, Y., Oemler, A., Lin, H., \& Tucker, D. L. 1998, \apj, 499, 589
%
\bibitem[Kennicutt (1998)]{ken98} Kennicutt, R. C. 1998, \araa, 36, in press
%
\bibitem[Landolt 1983]{lan83}
Landolt, A.U. 1983, \aj, 88, 439
%
\bibitem[Landolt 1992]{lan92} \underline{~~~~~~~} 1992, \aj, 104, 340
%
\bibitem[Lavery, Pierce \& McClure (1992)]{l92}
Lavery, R. J., Pierce, M. J., \& McClure, R. D. 1992, \aj, 104, 2067
%
\bibitem[Ledlow \& Owen 1995]{lo95}
Ledlow, M. J. \& Owen, F. N. 1995, \aj, 110, 1959
%
\bibitem[Morrison 1999]{m98}
Morrison, G. E. 1999, PhD Thesis, University of New Mexico
%
\bibitem[Newberry, Kirshner, \& Boroson (1988)]{new88}
Newberry, M. V., Kirshner, R. P., \& Boroson, T. A. 1988, \apj, 335, 629
%
\bibitem[Owen 1996]{ow96}
Owen, F. N. 1996, in {\it Extragalactic Radio Sources,
IAU 175}, Kluwer, (Ekers, Fanti, \& Padrielli, eds.), p305
%
\bibitem[Owen et al.,\ 1996]{oetal96}
Owen, F. N., Dwarakanath, K. S., Smith, C., Ledlow, M. J.,
Keel, W. C., Morrison, G. E., Voges, W., \& Burns, J., 1996
in {\it Energy Transport in
Radio Galaxies and Quasars}, (P. Hardee, A. Bridle, \& A. Zensus, eds.),
ASP Conference Series, Vol. 100, p. 353
%
\bibitem[Owen, Ledlow \& Keel 1995]{olk95}
Owen, F.N., Ledlow, M.J., \& Keel, W.C. 1995, \aj, 109, 14
%
\bibitem[Owen \& Laing 1989]{owl89}
Owen, F.N. \& Laing, R.A. 1989, \mnras, 238, 357
%
\bibitem[Roettiger, Burns, \& Loken 1993]{r93}
Roettiger, K., Burns, J. \& Loken, C. 1993. \apjl, 407, L53
%
\bibitem[Roettiger, Stone, \& Mushotzky 1998]{r98}
Roettiger, K., Stone, J. M., \& Mushotsky, R. F. 1998, \apj, 493, 62
%
\bibitem[Wang, Connolly, \& Brunner 1998]{w98}
Wang, Q. D., Connolly, \& Brunner, R. J. 1997, \apjl, 487, L13
%
\bibitem[Wilson 1996]{wil96}
Wilson, A.S. 1996, in {\it Energy Transport in
Radio Galaxies and Quasars}, (P. Hardee, A. Bridle, \& A. Zensus, eds.),
ASP Conference Series, Vol. 100, p. 9
\end{thebibliography}
\end{document}